\documentclass[useAMS,usenatbib]{mn2e}
\usepackage{psfig}
\begin{document}
\title[The HCO$^+$ Emission Excess in Bipolar Outflows]
{The HCO$^+$ Emission Excess in Bipolar Outflows}
\author[Rawlings et al]
{J.M.C Rawlings$^{1}$, M.P. Redman$^1$, E. Keto$^2$ and D.A. Williams$^1$\\
$^1$Department~of~Physics~and~Astronomy,
University~College London, Gower Street,London WC1E 6BT, UK\\
$^2$Harvard-Smithsonian Center for Astrophysics, 60 Garden Street,
Cambridge MA 02138, USA}
\date{\today}
\pubyear{2003}
\volume{000}
\pagerange{\pageref{firstpage}--\pageref{lastpage}}
\maketitle
\label{firstpage}

\begin{abstract}
A plausible model is proposed for the enhancement of the abundance of
molecular species in bipolar outflow sources.  In this model, levels
of HCO$^+$ enhancement are considered based on previous chemical
calculations, that are assumed to result from shock-induced desorption
and photoprocessing of dust grain ice mantles in the boundary layer
between the outflow jet and the surrounding envelope. A radiative transfer
simulation that incorporates chemical variations within the flow shows
that the proposed abundance enhancements in the boundary layer are
capable of reproducing the observed characteristics of the outflow
seen in HCO$^+$ emission in the star forming core L1527. The radiative
transfer simulation also shows that the emission lines from the
enhanced molecular species that trace the boundary layer of the
outflow exhibit complex line profiles indicating that detailed
spatial maps of the line profiles are essential in any attempt to
identify the kinematics of potential infall/outflow sources.  This
study is one of the first applications of a full three dimensional
radiative transfer code which incorporates chemical variations within
the source.

\end{abstract}


\section{Introduction}

The complex morphologies of molecular outflows have been identified in
numerous studies, mostly based on high resolution, interferometric CO
and optical surveys (eg. Arce \& Goodman, 2001, 2002; Lee et al. 2000,
2002). These surveys show that the molecular distributions are
generally composed of a large scale, poorly collimated low velocity
outflow that essentially traces the interaction between a surrounding
molecular cloud envelope and a collimated high velocity jet.  These
morphologies are hard to interpret unambiguously but often resemble
limb-brightened shells or sheaths surrounding cavities.

Observations at single-dish resolution suggest that the abundance of
HCO$^+$ in star forming cores with bipolar outflows may be enhanced by
a factor of up to 100--1000$\times$ relative to dark-cloud values -
for example, a fractional abundance of $4\times 10^{-8}$ is implied in
the case of the Class 0 source B335 (Rawlings, Taylor and Williams
2000).  This, and similar observations, provided the basis for the
models of Rawlings~et~al.\@~(2000) in which the
enhancement of HCO$^+$ originates in the turbulent mixing layer which
is the interface between the high velocity outflow and the quiescent
or infalling core gas which it is steadily eroding. The molecular
enrichment is driven by the desorption of molecular-rich ice mantles,
followed by photochemical processing by shock-generated radiation
fields.  Thus, CO and H$_2$O are released following mantle
evaporation.  The CO is then photodissociated and the C that is
produced is photoionized by the shock-induced radiation field that is
generated in the interface:

\[ {\rm CO} + {\rm h}\nu \longrightarrow {\rm C} + {\rm O} \]
\[ {\rm C} + {\rm h}\nu \longrightarrow {\rm C}^+ + {\rm e}^- \]
The subsequent reactions of C$^+$ with H$_2$O lead to an enhancement of the
HCO$^+$ abundance:
\[ {\rm C}^+ + {\rm H}_2{\rm O} \longrightarrow {\rm HCO}^+ + {\rm H} \]
This abundance enhancement is only temporary, and progresses as a wave that
fans out from the interface into the core. Such behaviour is both qualitatively
and quantitatively consistent with the scenario proposed by Velusamy and Langer
(1998) for B5 IRS1 on the basis of their $^{12}$CO~(2-1) and C$^{18}$O~(2-1)
observations; the data clearly show a parabolic outflow cavity that appears to
be steadily widening, so that the opening angle is growing at a rate of
0.006$^\circ$yr$^{-1}$.
An extension of the chemistry to a larger species/reaction set by Viti,
Natarajan and Williams (2002) confirmed this result and concluded that other
molecular tracers, such as H$_2$CS, SO, SO$_2$ and CH$_3$OH could be similarly
affected.

At higher resolutions, aperture synthesis observations are capable of
identifying the morphology of the interaction between the jets and
their surroundings.  Hogerheijde et al. (1998) made high resolution
HCO$^+$ J=1$\to$0 observations using the Owens Valley Millimetre Array
(OVRO) towards a number of low mass YSOs. One of the most interesting
results from that survey was the detection of compact emission
associated with the walls of an outflow cavity in L1527 (Hogerheijde
et al. 1998, figure 1). The $^{12}{\rm CO} 3-2$ outflow lobes are
well-developed, with lengths of $\sim 0.12~\rm pc$ and a large opening
angle ($\sim 90^{\circ}$). The single dish observations of Hogerheijde
et al (1997) show that the HCO$^+$ emission is confined to a smaller
scale, $\sim 0.015$ to 0.022~pc, in the center of the flow. The
interferometric observations with higher angular resolution show that
much of this HCO$^+$ emission originates in the boundary of the lobes
of the CO outflow rather than in the quiescent core of the cloud
surrounding the outflow.  The enhanced HCO$^+$ emission exhibits a
cross-shaped morphology which is, of course, undetectable at
single-dish telescope resolution (e.g. see Hogerheijde et al. 1997,
figure 3). The data is also consistent with an HCO$^+$ enhancement by
a factor of about 10$\times$ in the outflow sheaths.

The region of apparent HCO$^+$ emission excess has a limited extent,
$20-30\arcsec\sim 3000-4500~{\rm AU}$, at an assumed distance of
150~pc, a factor of 10 smaller than the CO emission. If the HCO$^+$
were enhanced over the length of the outflow, then more extended
HCO$^+$ emission would be expected in the single-dish observations
even if the enhanced region were at a scale too large to be detected
by the interferometer (the OVRO will not recover any emission larger
than $\sim 30\arcsec$). The small extent of the HCO$^+$ emission
implies that the enhancement is temporary.  For example, if the
HCO$^+$ molecules created in the boundary layer of the outflow close
in to the star are not destroyed, then the molecules ought to be
transported down the outflow and ought to be detected in the single
dish observations. The extent of the HCO$^+$ provides a constraint on
the timescale for decay. The position-velocity maps for this source
(Hogerheijde et al. 1998, figure 7) suggest that the outflow velocity
in the emission-enhanced mixing layer is only $\sim 5~{\rm
km~s^{-1}}$.  This implies a dynamical age for the HCO$^+$ emitting
gas of just $\sim$3100 yrs, assuming the outflow is close to the plane
of the sky.  We therefore require a mechanism to restrict the HCO$^+$
enhancement to this region with a timescale of just a few thousand
years.

As stated above, this source has an outflow with a wide opening angle
($\sim 90^{\circ}$).  Note that the $^{12}\rm CO$ emission is much
more extensive and has a larger dynamical age than the HCO$^+$
emission excess gas. The orientation of the outflow axis is close to
the plane of the sky for this source which implies that the outflow
and core emission will be not be clearly separable.  It should also be
noted that the dense core may be comparable to this size which could
truncate the HCO$^+$ emission.

However, there is still some ambiguity as to the interpretation of the
morphology; whether it genuinely traces the abundance, temperature or
density enhancements in the outflow interface, or whether it is a
simply an excitation effect; the elevation of the HCO$^+$ excitation
temperature in the boundary layer perhaps deriving from the outflow
cavity acting as a low opacity pathway for photons from the star-disk
boundary layer (Spaans et al. 1995).  However, we must also recognise
that it would be most unlikely for the HCO$^+$ abundance in the mixing
layer to be the same as in the surrounding core.  Previous models of
boundary layers, including both analytical studies of turbulent
interfaces (e.g. Rawlings and Hartquist 1997) and numerical
hydrodynamical calculations (Lim, Rawlings and Williams 2001) have all
shown strong HCO$^+$ abundance enhancements within the
interface. Moreover, these studies have not included the effects of
gas-grain interactions which are likely to further enhance the HCO$^+$
abundance.

In this paper we do not attempt to model the chemical processes within
the interface in any detail. Rather we consider the possible levels of
molecular enhancement that are consistent with the observations.  This
is a timely precursor to the more detailed chemical/radiative transfer
calculations that will be required as more higher resolution
observational facilities become available. These facilities will be
capable of resolving the dynamics and internal structures of sources
on milliarcsecond scales.

We have considered two different scenarios that can result in
molecular enhancement (and specifically, of HCO$^+$) in the outflow
sheath which forms the interface between the bipolar outflow jet and
the molecular core envelope:
\begin{itemize}
\item Molecular enrichment through entrainment of dense circumstellar (disk)
material into the outflow, and
\item Shock-induced mantle sublimation and photochemical enhancements (as in
Rawlings~et~al.\@~2000), dependent on the precise geometry of the outflow.
\end{itemize}

\section{The Radiative Transfer Code}

In order to convert the results from a chemical model to images of
molecular line brightnesses, line profile shapes and velocity channel
maps it is necessary to employ a molecular radiation transport
code. This is particularly true when considering deviations from
spherical symmetry and observations at high resolution.

An important failing of previous studies (e.g. Rawlings, Taylor and
Williams, 1998; Viti, Natarajan and Williams, 2002) is that they could
only make bulk predictions relating to line of sight abundance
enhancement. The hypotheses stated in those papers could not be
verified because there was no way at the time of checking the
consistency of the predictions through comparison with line profile
shapes, intensity maps etc.

In order to remedy this situation, we have used the radiative transfer
code described in Keto et al. (2003). In dark molecular clouds, the
density is too low for local thermodynamic equilibrium (LTE) to apply,
the opacity is too high for an optically thin approximation and the
systemic velocities are too low for the large velocity gradient (LVG)
or Sobolev approximations to be valid.  We therefore generate model
spectra and intensity maps using a non-LTE numerical radiative
transfer code. The code is fully 3 dimensional and employs the
accelerated lambda iteration (ALI) algorithm of Rybicki and Hummer
(1991) to solve the molecular line transport problem within an object
of arbitrary (three-dimensional) geometry.  This represents a
considerable improvement over the early models of Keto (1990) - as
applied (for example) to the modelling of the accretion flow onto a
star-forming core in W51 (Young, Keto and Ho 1998). These codes
assumed a simple determination of the level populations using either
the LTE or the Sobolev approximation, together with iterative
correction for the effects of the coupling between multiple resonant
points in the source (Marti and Noerdlinger 1977; Rybicki and Hummer
1978).

The local line profile is specified by systemic line-of-sight motions
together with thermal and turbulent broadening. In addition to
essential molecular data (such as collisional excitation rates) the
physical input to the model consists of the three-dimensional
velocity, density and temperature structures. The radiative transfer
equations are then numerically integrated over a grid of points
representing sky positions. The main source of uncertainty in the
radiative transfer calculations are the collisional rate coefficients,
but this should only be manifest in the uncertainties in the line
strengths (which may be inaccurate by as much as 25\%); the line
profile shapes and and relative strengths should be less affected. The
${\rm HCO^+}$ collision rates of Flower (1999) are used.

For our purposes we define a spherical cloud within a regularly spaced
cartesian grid of $96\times 96\times 96$ cells. Within each cell, the
temperature, ${\rm H}_2$ density, molecular species abundance,
turbulent velocity width, and bulk velocity are specified. Externally,
the ambient radiation field is taken to be the cosmic microwave
background. At a given viewing angle to the grid, the line profile is
calculated at specified offsets from the centre by integrating the
emission along the lines of sight.  It is then possible to regrid the
data in the map plane and convolve with the (typically Gaussian) beam
pattern of the telescope. The spectra are computed with a frequency
resolution of $0.01~{\rm km~s^{-1}}$ and Hanning smoothed to
$0.02~{\rm km~s^{-1}}$. 

\section{Modelling an outflow}

As our test object against which to compare the models we use the
observational data from Hogerheijde et al. (1998) for L1527. In this
source the orientation of the outflow, close to the plane of the sky,
is favourable for locating the boundary layers of the outflow, seen in
the interferometric observation of HCO$^+$, with respect to the more
extended outflow seen in CO.  In addition, the availability of both
single dish and interferometric observations of HCO$^+$ helps separate
the HCO$^+$ emission in the outflow boundary layer from any emission in
the quiescent core.  The interferometric observations tend to resolve
out the emission from the larger scale core while emphasising the
emission from smaller scale boundary layer. The single dish
observations include emission from both the large and small scale
structures averaged together within the lower angular resolution
single dish beam.

\subsection{Disk matter injection}

In this model, rather than the injection of the HCO$^+$ occurring
continuously along the outflow, we assume that molecular-rich gas is
injected into the outflow as a result of its interaction with a cool,
dense, remnant protostellar accretion disk.  It is then transported
away within the mixing layer.  This molecular entrainment would occur
very close to the base of the jet, on scales that are essentially
unobservable. The fact that the region of molecular enhancement has a
limited size could thus be interpreted simply as a consequence of the
one-point injection of material; molecules are injected at the origin
and they decay as the flow progresses downstream.

The chemical evolution of the parcel of gas is followed using a simple
one-point variation of the models of Rawlings~et~al.\@~(2000), using
the same physical assumptions and chemical initial conditions.  Thus,
the chemistry includes some 1626 reactions between 113 species. The
initial chemical enrichment is calculated on the assumption that prior
to shock activity gas-phase material froze out onto the surfaces of
dust grains and was rapidly hydrogenated. At the start of the
calculations (i.e. in the immediate vicinity of the protostar) we
assume that the molecular-rich (H$_2$O, CO, CH$_4$, NH$_3$) mantles
are shock-sputtered and immediately released back into the gas-phase
and we then follow the subsequent photochemical/kinetic evolution
downstream.

Using the model of Rawlings~et~al.\@~(2000) as the basis set we
considered three sets of parameters: \\
\underline{\em Model 1a:} the standard calculation, in which the total hydrogen
density, n$_H$=10$^5$cm$^{-3}$, the temperature T=50K and the ultraviolet flux
F$_{UV}$(total)=10$\times$F$_{UV}$(ISM),\\
\underline{\em Model 1b:} the same as Model 1a, but with a lower radiation
field strength: F$_{UV}$(total)=F$_{UV}$(ISM), and\\
\underline{\em Model 1c:} the same as Model 1b, but with a lower density
(n$_H$=10$^4$cm$^{-3}$) and higher temperature (T=100K) - values used as
standard parameters in Rawlings~et~al.\@~(2000) - but with the lower
radiation field.\\
In all models we assume that A$_v$=0.0 and the CO (line) dissociating radiation
field strength is equal in strength to the interstellar radiation field.

The time-dependence of the HCO$^+$ abundance from each of these models is shown
in table~\ref{injection}.
\begin{table}
\begin{tabular}{l|l|l|l}
\hline
Time (years) & Model 1a & Model 1b & Model 1c\\
\hline
10   & $2.7\times 10^{-8}$ & $1.2\times 10^{-7}$ & $3.2\times 10^{-7}$\\
20   & $2.4\times 10^{-9}$ & $1.0\times 10^{-7}$ & $3.0\times 10^{-7}$\\
50   & $3.7\times 10^{-12}$ & $5.7\times 10^{-8}$ & $1.7\times 10^{-7}$\\
100  & $2.5\times 10^{-12}$ & $2.0\times 10^{-8}$ & $5.4\times 10^{-8}$\\
200  & $1.9\times 10^{-12}$ & $1.2\times 10^{-10}$ & $3.9\times 10^{-9}$\\
500  & $1.1\times 10^{-12}$ & $2.3\times 10^{-12}$ & $1.9\times 10^{-11}$\\
1000 & $1.0\times 10^{-12}$ & $1.8\times 10^{-12}$ & $5.9\times 10^{-12}$\\
\hline
\end{tabular}
\caption{The time-dependence of the HCO$^+$ fractional abundance in models
where molecular material is injected at the base of the outflow}
\label{injection}
\end{table}

Inspection of the results reveals that in all three scenarios, the
abundances rapidly decay from X(HCO$^+$)$\sim 10^{-7}$ to less than
X(HCO$^+$)$\sim 10^{-9}$ within just a few hundred years at most. The
dominant destruction pathway is dissociative recombination
(Rawlings~et~al.\@~2000).  With a jet speed of $\sim 5~{\rm
km~s^{-1}}$, this corresponds to a distance of just $\sim
10^{15}-10^{16}~{\rm cm}$. This is in contradiction to the
observations of Hogerheijde et al which show lobes of HCO$^+$ emission
extending significantly into the larger CO outflow.  We may therefore
conclude that this mechanism cannot explain the observed HCO$^+$
enhancements in the outflow, which must therefore be generated locally
in the interface.

\subsection{Geometry-dependent molecular enrichment}

In this model we assume that HCO$^+$ is formed (as in the
Rawlings~et~al.\@~2000 model) as a result of shock interaction between
the jet and its surroundings, and the HCO$^+$ is continuously
generated at the boundary layer of the flow.  On the simple assumption
that the shock-induced radiation field, and hence the photochemistry,
is stronger in regions of greater shock activity, the level of the
HCO$^+$ enhancement should thus be directly related to the the degree
of shock activity which in turn might depend on the angle of
interaction between the jet and the envelope gas.  We can model this
process by including an HCO$^+$ source term that varies with the local
{\it curvature} of the interface of the outflow. The hope is that this
simple model would then be capable of explaining why the HCO$^+$
emission morphology has the shape, dimensions and contrasts that are
observed.  Obviously, this model is very simplistic in its nature and
does not address the many uncertainties in the free-parameters, such
as the shock strengths and spectrum, the radiation fields or a
detailed consideration of the molecular injection rates, which may all
be very complex functions of position along the flow. However, from
the previous section it is apparent that the HCO$^+$ enhancement has
to be generated locally, but that local enhancement has a very limited
spatial extent. We have therefore postulated this very simple
empirically-based model.

Three distinct zones were specified in order to model the system:
\begin{itemize}
\item a quiescent envelope
\item a jet
\item a boundary layer between the jet and the envelope
\end{itemize}
In our first simple models, the jet is assumed to be of high velocity
and high temperature but low density whilst the core envelope has a
low ($\sim 0$) velocity and temperature and high density. The boundary
layer is the interface between the outflow and the envelope and is a
hot mixing layer with high velocity and temperature. We assume that
the trace molecule HCO$^+$ is injected along the interface layer,
presumably by the shock-induced sputtering/photochemical mechanism
described by Rawlings~et~al.\@~(2000). We have considered
two possible distributions for the HCO$^+$ fractional abundance
[X(HCO$^+$)];
\begin{enumerate}
\item X(HCO$^+$)$=$constant at all positions along the boundary layer, and
\item X(HCO$^+$) is proportional to the inverse of the radius of curvature.
As discussed in the previous section, X(HCO$^+$) declines rapidly
following its injection into the boundary layer. Thus X(HCO$^+$) may
be expected to track the injection rate as a function of position
along the flow. As explained above, it would seem plausible to suggest
that the HCO$^+$ injection rate and hence its abundance are simply
related to the radius of curvature of the boundary layer.
\end{enumerate}

In all of our models we adopt a very simple physical representation of
the boundary layer, so that at any radial point along the interface,
the density, temperature and velocity are all constant across the
spatial extent of the interface. Thus, we do not include an allowance
for shear velocity structure, nor temperature and density variations
between the jet edge and the envelope edge of the mixing layer. Note
that although the employed source models are 2D in nature, the 3D
functionality of the code is used.

Two simple geometries were investigated for the shape of the outflow;
\begin{itemize}
\item[(a)] A wide-angled conical outflow
\item[(b)] An `hour-glass' shaped outflow defined by a tanh function
\end{itemize}
For (a), the conical outflow has a constant opening angle of
$45^\circ$. The boundary layer was specified as being the region
within $\pm 7.5^\circ$ of this angle. In this model we assumed that
X(HCO$^+$)=0 in the jet/cavity, whilst in the boundary layer it is
enhanced by a factor of 10$\times$ over the dark cloud/envelope value,
but does not vary with position along the boundary layer. We also
assume that since the boundary layer has a constant opening angle, its
density is subject to spherical dilution and hence varies as
$r^{-2}$. The temperature is assumed to be constant at all points
along the boundary layer.
\begin{table*}
\begin{tabular}{l|l|l|l}
\hline
~~ & Envelope & Jet/Cavity & Boundary Layer \\
\hline
Radial velocity $({\rm km~s^{-1}})$ & 0.05 & 5.0 & 5.0 \\
Density (${\rm cm^{-3}}$) & $1.0\times 10^{4}$ & $1.0\times 10^{3}$ & $1.0\times
10^{4}$\\
X(HCO$^+$)~$\times 10^{-10}$ & 1 & 10 & 1000 \\
Temperature (K) & 10 & 50 & 50 \\
Microturbulent velocity $({\rm km~s^{-1}})$ & 0.2 & 1.0 & 1.0 \\
\hline
\end{tabular}
\caption{Physical parameters used in the best fit boundary layer model}
\label{parameters}
\end{table*}

For (b), a {\em tanh} function, with a scaling parameter, was used to
mimic the typical hourglass shape that is seen in many outflows. The
equation of the inside and outside edges of the boundary layer are,
respectively
\begin{equation}
z=\alpha_{\rm in}\tanh(\alpha_{\rm in}\lambda r),
\end{equation}
\begin{equation}
z=\alpha_{\rm out}\tanh(\alpha_{\rm out}\lambda r).
\end{equation}
Thus, $\lambda$ defines the shape of the flow and $\alpha_{\rm in}$,
$\alpha_{\rm out}$ define the edges (and hence the thickness) of the
boundary layer. The jet was specified as having a constant radial
velocity. Within the boundary layer, the velocity of the gas was
maintained at a constant speed $V_{\rm b}$ with a direction tangential
to the inside edge of the boundary layer. Thus the velocity components
in terms of cylindrical coordinates, $r$ and $z$ are
\begin{equation}
v_{r} =\frac{V_{\rm b}\lambda(1-r^2)}{\left[1+\lambda^2(1-r^2)^2\right]^{1/2}},
\end{equation}
\begin{equation}
v_{z} =\frac{V_{\rm b}}{\left[1+\lambda^2(1-r^2)^2\right]^{1/2}},
\end{equation}
For the sake of simplicity, in this model the density is taken to be constant 
within the boundary layer.

For the models where X(HCO$^+$) depends on the local radius of curvature (ii), 
this implies that the abundance then varies as
\begin{equation}
{\rm X(HCO^+)}\propto {\rm sech}^2(\lambda z)~{\rm tanh}(\lambda z)
\end{equation}
normalised to some empirically constrained maximum value at the most curved 
part of the flow. 

The core envelope radius is taken to be 0.3pc, and other parameters
for the envelope, jet/cavity and boundary layer are given in
Table~\ref{parameters}. This table gives the values of the various
free parameters that give the best fit to the observations.  Although
values are given for the radial velocity, HCO$^+$ abundance,
temperature and microturbulent velocity in the jet/cavity region, the
density is very much lower in the cavity than in the envelope and
boundary layer so that it contributes very little to the HCO$^+$ line
emission.  The results are therefore very insensitive to these values.

\section{Results}

Adopting the physical parameters in Table~\ref{parameters}, the only
free physical parameters in the conical outflow models are the opening
angle of the outflow ($\theta$) and the thickness, expressed as an
opening angle ($\Delta\theta$), of the mixing layer.  In the case of
the {\em tanh} outflow models, the only free parameters are the geometry
parameter ($\lambda$) and the boundary layer limits ($\alpha_{\rm in}$,
$\alpha_{\rm out}$).  We have considered values of $\theta =45^{\circ}$,
$\Delta\theta =$ 10, 15 and 20$^{\circ}$ for the conical model, and
investigated $\lambda=$0.5, 1.0, 2.0, 3.0 for a range of values of
$\alpha_{\rm in}$, $\alpha_{\rm out}$ for the 'tanh' model. The closest fit to
the observed morphology and intensity contrast in L1527 was obtained
with $\lambda=3.0$, and $\alpha_{in}$, $\alpha_{out}$=1.0 and 1.1
respectively. This is the `best fit' model whose results are
illustrated and discussed below.

Maps of the integrated intensity, line profiles and intensity contour
maps have been calculated for three different inclination angles of
the outflow relative to the observer: $\phi=0^{\circ}$ (which
corresponds to the outflow being in the plane of the sky),
$\phi=60^{\circ}$, and $\phi=90^{\circ}$ (which corresponds to the
outflow being observed pole-on).  Results have been calculated for the
three observable low lying transitions of HCO$^+$: J=1$\to$0, 3$\to$2
and 4$\to$3. The emission modeled by the radiative transfer code is
convolved with a $1.5\arcsec $ Gaussian beam. This is a smaller beamsize
than the $5\arcsec$ OVRO observations described by Hogerheijde et al
(1998) and is used to allow fine detail in the numerical model to be
discerned.

In general, regardless of the assumed morphology of the outflow, we
find that if the boundary layer is too thick, all lines of sight that
pass through the outflow cavity intercept regions of optically thick
emission and thus the morphology resembles a filled cross or `bow
tie'. For example, in the {\em tanh} model with $\lambda=3.0$, and
$\alpha_{\rm in}=1.0$ the model morphology begins to become a poor fit
to the observations for $\alpha_{\rm out}\ga 1.2$. This implies a
rough upper limit to the boundary layer thickness of $\sim 400~{\rm
AU}$

The results for the conical jet model (a) are as expected - the
observed HCO$^+$ emission excess can be modelled but, even allowing
for an inverse square law for the fall in density with radial
distance, the HCO$^+$ emission excess is spatially extended and we
cannot reproduce the observed compact structures which have fairly
sharp cut-offs in emission beyond about 0.02pc.

However, using the values of the model parameters given in
Table~\ref{parameters} together with the geometry parameters for the
`best fit' model for a {\em tanh}-shaped outflow [model (b)], results
in a very close fit to the observed HCO$^+$ morphology of L1527.
Smaller (larger) values of $\lambda$ give wider (narrower) opening
angles to the jet. The observational consequences are variations in
both the observed thickness of the limb-brightened wings and the
intensity contrast between the wings and the cavity lines of
sight. Models in which the HCO$^+$ abundance does not vary with
position [(i)] are incapable of matching the observed emission
morphology regardless of the choice of values for the other free
parameters.  The best fit model therefore has the HCO$^+$ abundance
variation as described in (ii), above, and normalised so that peak
boundary layer abundance is that given in
Table~\ref{parameters}. These values are comparable to those of
Rawlings~et~al.\@(2000).

\begin{figure*}
\psfig{file=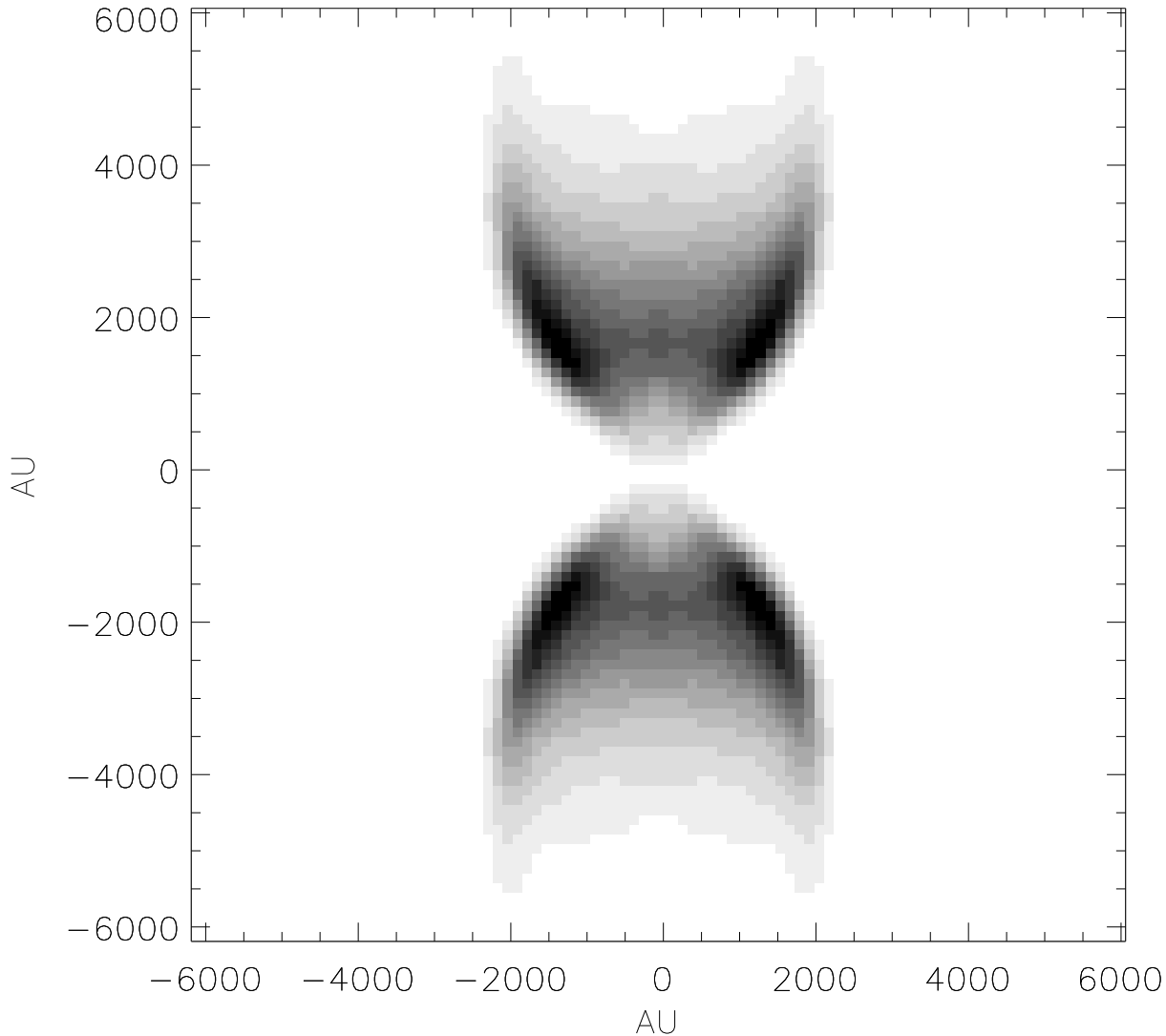,width=450pt,bbllx=54pt,bblly=251pt,bburx=400pt,bbury=566pt}
\caption{Linear greyscale representation of HCO$^+~(1-0)$ best model. The peak of the emission is $5.2~{\rm K~km~s^{-1}}$. The viewing angle is $0\deg$}
\label{plot1}
\end{figure*}
\begin{figure*}
\psfig{file=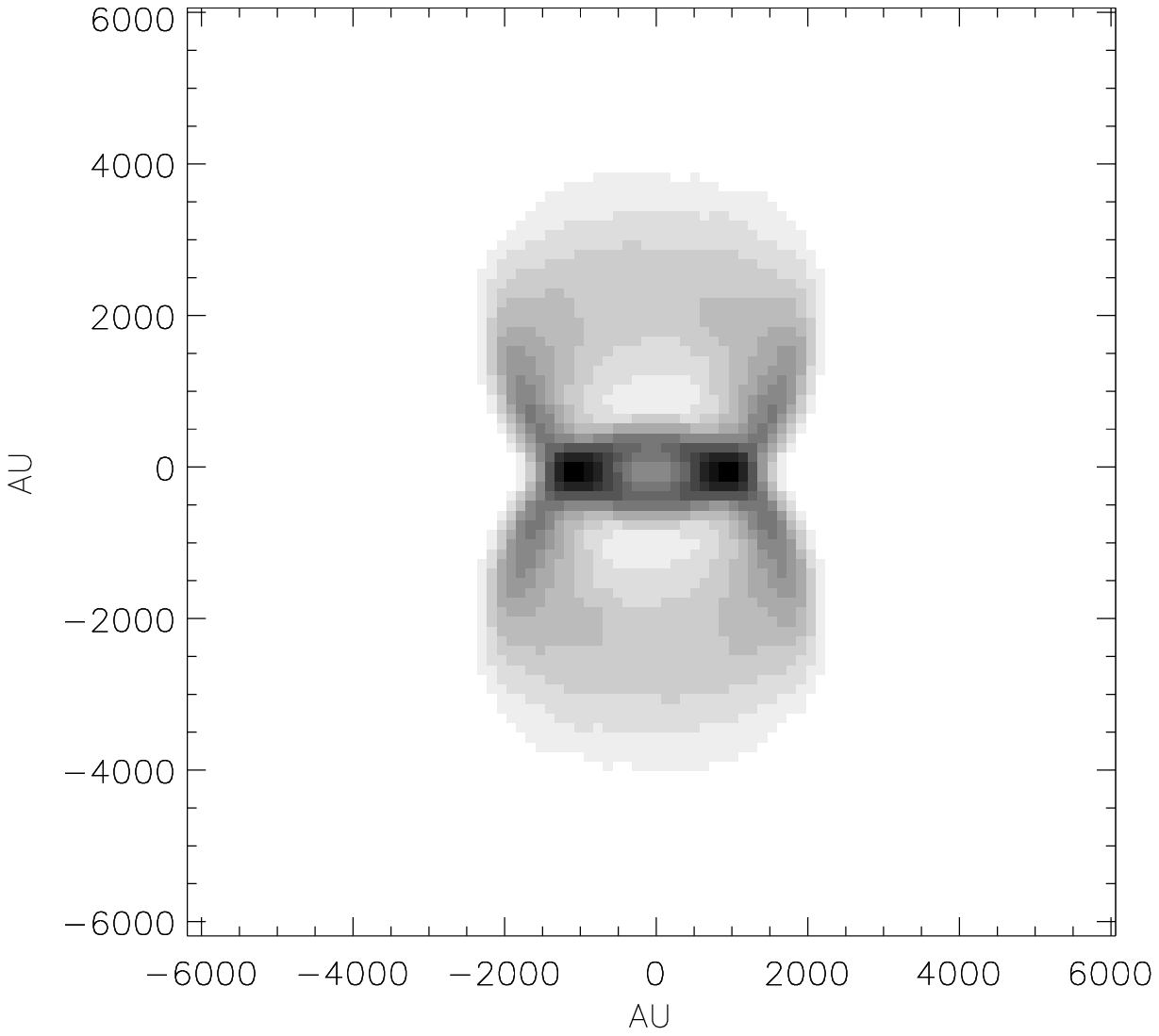,width=450pt,bbllx=54pt,bblly=251pt,bburx=400pt,bbury=566pt}
\caption{Linear greyscale representation of HCO$^+~(1-0)$ best model. The peak of the emission is $7.0~{\rm K~km~s^{-1}}$. The viewing angle is $60\deg$}
\label{plot2}
\end{figure*}

For the best fit model we have also considered the possibility of
temperature variations along the boundary layer. We assumed that the
temperature varies in a similar manner to the HCO$^+$ abundance
variation in (ii). This would seem plausible if both the HCO$^+$
enhancement and the excess heating arise from shock activity. Thus, we
again made the assumption that the shock activity and hence
temperature is inversely proportional to the radius of curvature, and
is normalised so that it lies in the range 10--100K. The effect on the
overall morphology of the emission is much less marked than the effect
of varying the HCO$^+$ abundance and does not even have very strong
effects of the 1$\to 0: 3\to 2: 4\to 3$ line ratios. Therefore, to
simplify the analysis, the best fit model whose results we present
here is one in which the temperature is kept constant along the
boundary layer.
\begin{figure}
%
\psfig{file=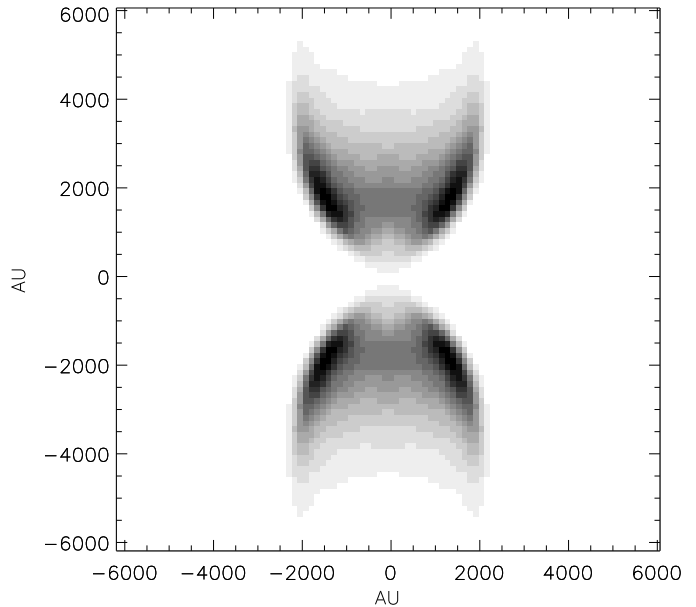,width=250pt,bbllx=54pt,bblly=251pt,bburx=400pt,bbury=566pt}
\caption{Linear greyscale representation of HCO$^+~(3-2)$ best model. The peak of the emission is $0.74~{\rm K~km~s^{-1}}$. The 
viewing angle is $0\deg$}
\label{plot3}
\end{figure}
\begin{figure}
%
\psfig{file=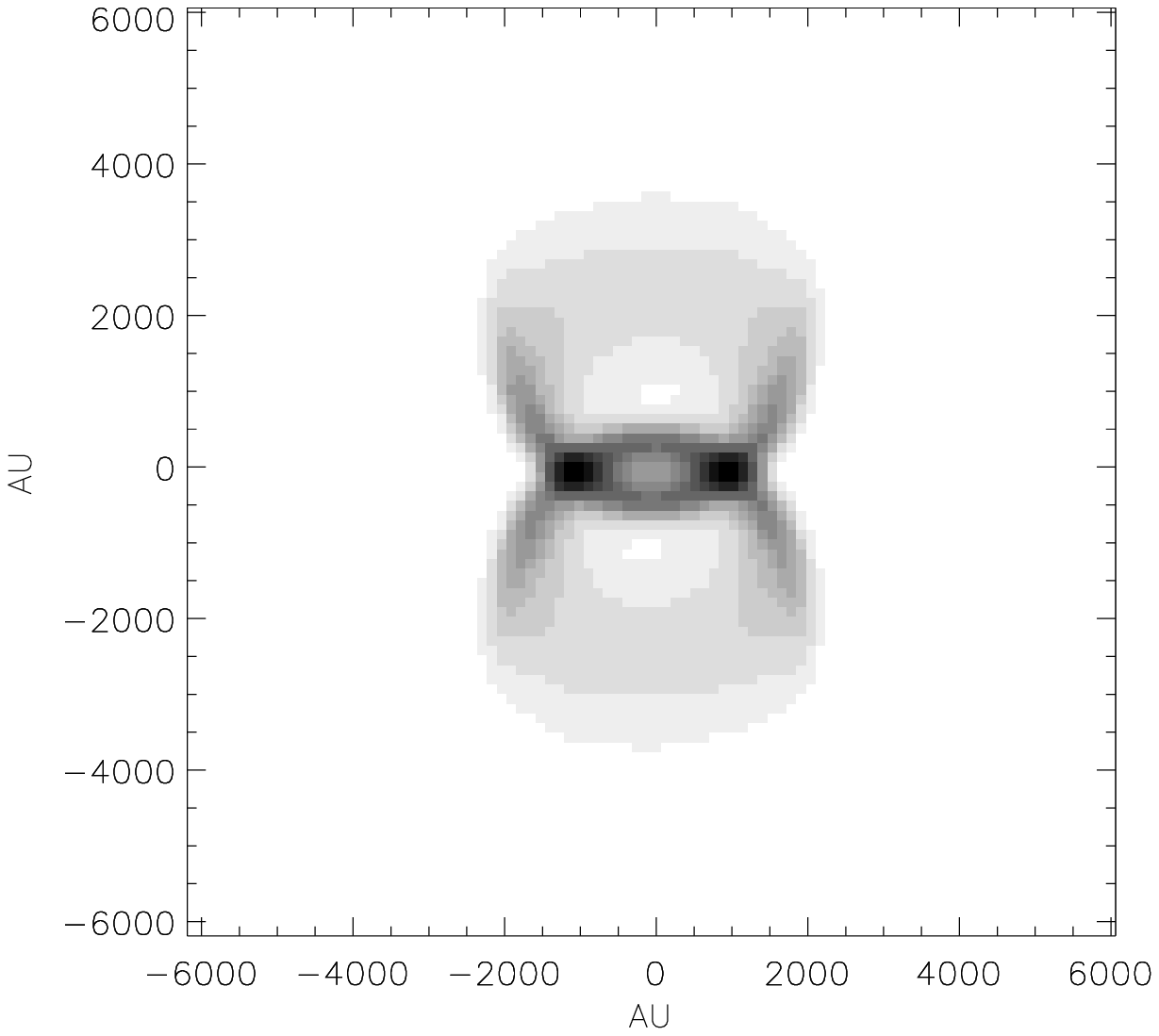,width=250pt,bbllx=54pt,bblly=251pt,bburx=400pt,bbury=566pt}
\caption{Linear greyscale representation of HCO$^+~(3-2)$ best model. The peak of the emission is $1.3~{\rm K~km~s^{-1}}$. The viewing angle is $60\deg$}
\label{plot4}
\end{figure}

\begin{figure*}
\psfig{file=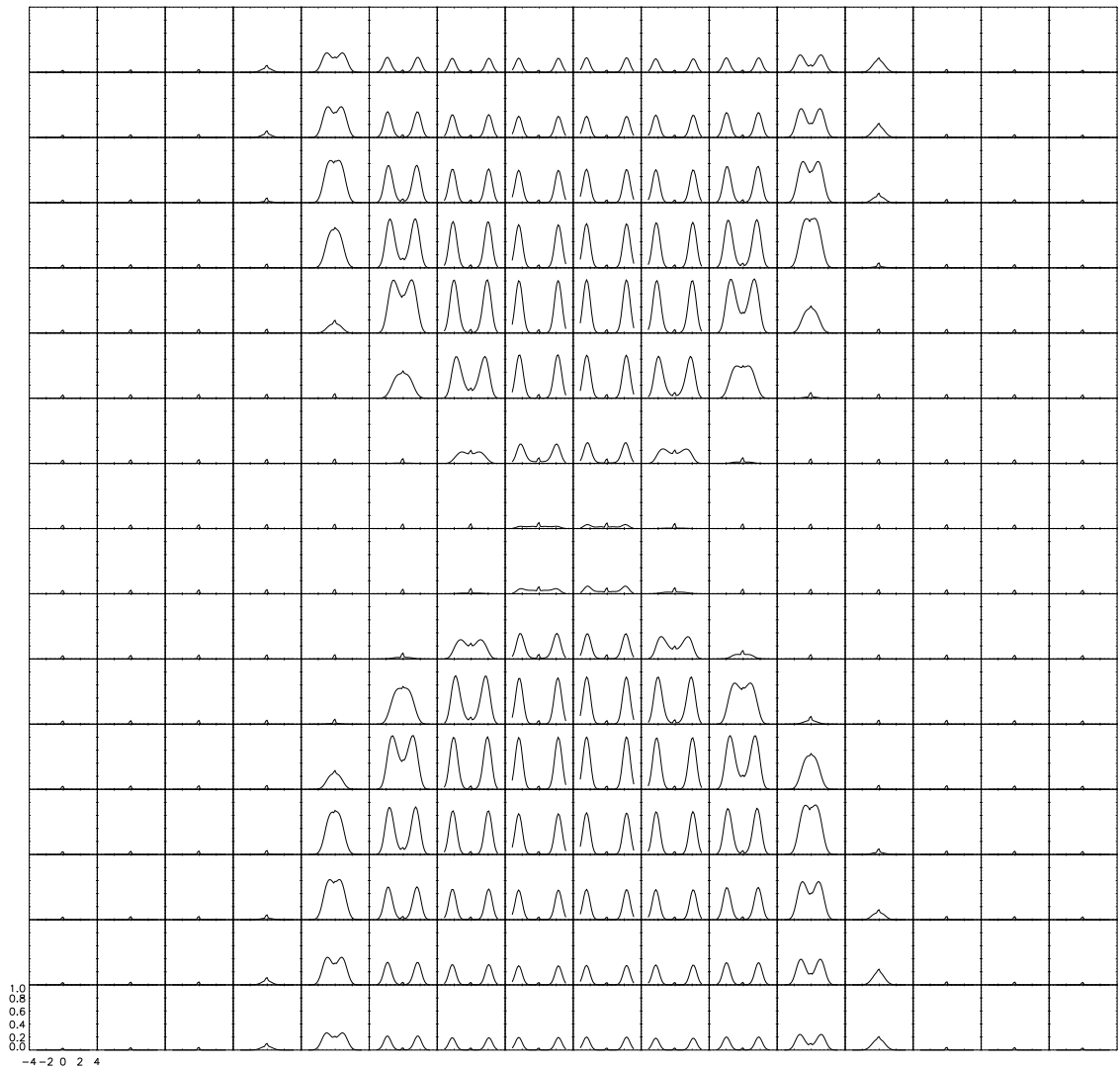,width=450pt,bbllx=0pt,bblly=0pt,bburx=581pt,bbury=544pt}
\caption{
Line profiles of HCO$^+~(1-0)$ for the best model. For clarity, a
subset of the 96x96 array of line profiles is displayed: only the
central regions are shown and only every fourth profile, in both X and
Y directions, is displayed. The viewing angle is $0\deg$. The tiny
peaks of emission are from the cold envelope gas}
\label{plot5}
\end{figure*}

\begin{figure*}
\psfig{file=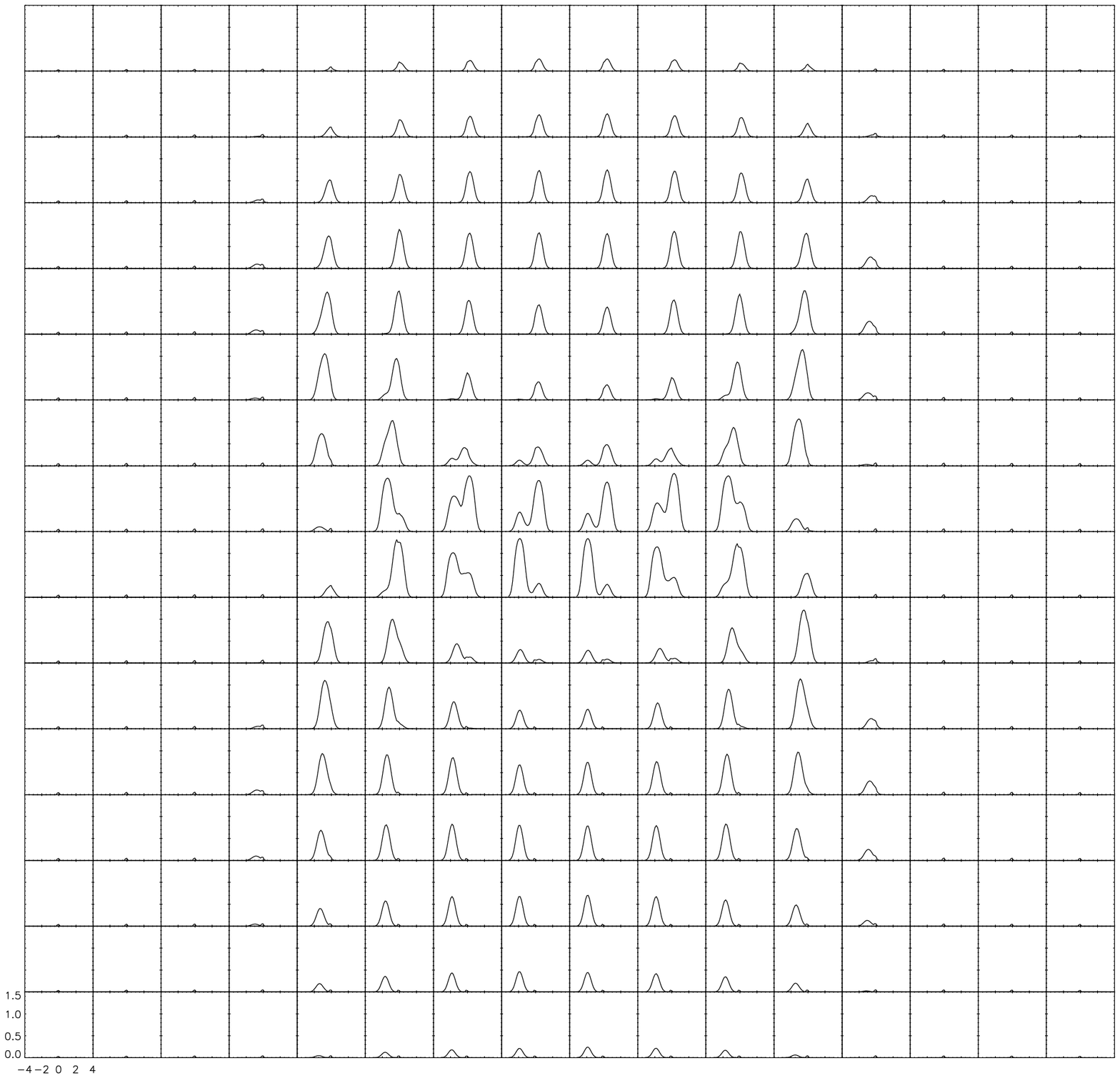,width=450pt,bbllx=0pt,bblly=0pt,bburx=616pt,bbury=542pt}
\caption{Line profiles of HCO$^+~(1-0)$ for the best model. For clarity, a
subset of the 96x96 array of line profiles is displayed: only the
central regions are shown and only every fourth profile, in both X and
Y directions, is displayed. The viewing angle is $60\deg$. The tiny
peaks of emission are from the cold envelope gas}
\label{plot6}
\end{figure*}

Figures~\ref{plot1} to \ref{plot6} all show results from the same
(`best fit') model, but for different transitions and viewing angles.
Figure~\ref{plot1} shows a grayscale representation of the integrated
line intensity for the HCO$^+$(1$\to$0) line. For this particular
calculation, the viewing angle is 0$^{\circ}$ (i.e. the outflow is in
the plane of the sky, so that the viewing angle is perpendicular to
the axis of the jet). Very similar results are obtained for the $3\to
2$ and $4\to 3$ transitions.  Note the lack of emission close to the
origin of the flow, consistent with the observations and essentially a
result of the fact that the curvature of the flow - and hence the
implied shock activity and chemical enrichment - vanish at the
origin. Note also the strongly limb-brightened emission from the edges
of the cavity. Again, the thickness, shape (including the location of
the peaks of the emission) and intensity contrast match the
observations well. 

The model emission compares very well with the observations, given the
uncertainties dues to the collision rates, as discussed in Section
2. Hogerheijde et al. (1998) measure an integrated intensity in the
HCO$^+$(1$\to$0) line of $< 2.0~{\rm K~km~s^{-1}}$ over their
$5\arcsec$ beam. The peak emission for the best fit model, as
displayed in Figure 1, is $5.4~{\rm K~km~s^{-1}}$ over a $1.5\arcsec$
beam and when this model is convolved to a $5\arcsec$ beam, the peak
emission becomes $3.2~{\rm K~km~s^{-1}}$ (since the emission is
concentrated in the bright boundary layer).

Figure~\ref{plot2} also shows a grayscale map of the integrated line
intensity for the $3\to 2$ line, but in this case for a tilted
(60$^{\circ}$) viewing angle. Figures~\ref{plot3} and \ref{plot4} show
greyscale maps of the integrated line intensity
for the HCO$^+$(3$\to 2$) line for different viewing angles (0$^{\circ}$ and
60$^{\circ}$).

Figure~\ref{plot5} gives a grid map of the spectral line profiles for
the HCO$^+$($1\to 0$) line for an assumed viewing angle of
0$^{\circ}$.  Again, similar results are obtained for both the ($3\to
2$) and the ($4\to 3$) line maps. At the extreme edges of the
cavity/boundary layer the lines of sight essentially intercept the
boundary layer cone at one point with a relatively long path length
through the boundary layer - hence the single peaked, limb-brightened
emission. Along other lines of sight (and most strongly along lines of
sight that intercept the outflow axis) the path intercepts the
boundary layer cone at two points; the near side (which is approaching
the observer) and the far side (which is receding) - hence the double
peaked emission. The separation of the peaks (and the difference
between the projected line of sight velocities of the two components)
falls as one moves away from the outflow axis.

Figure~\ref{plot6} gives a grid map of profiles for the HCO$^+$($1\to
0$) line and a viewing angle of 60$^{\circ}$. Note that the line
profiles are qualitatively quite different to those shown in
fig.~\ref{plot5}. With the outflow axis tilted out of the plane of the
sky, the whole of the lower outflow lobe is tilted towards the
observer, whilst the upper lobe is tilted away from the observer. As a
consequence the line profiles show a net redshift/blueshift in the
upper/lower lobes.

Most interestingly, there is some rather complex behaviour close to
the origin; the line profiles are asymmetric and self-reversed, but
along some lines of sight the red wing is stronger than the blue,
whilst along other, nearby, lines of sight the opposite is
true. Bearing in mind the inclination (60$^{\circ}$) of the source
this is an effect of observing parts of the flow that are receding and
parts that are approaching the observer along the line of
sight. Looking near the centre of the bipolar flow, the line of sight
intersects more than one surface of the bipolar cones. This
observation does, however, emphasize that it is very difficult to
interpret single line profiles in the context of inflows/outflows
without a clear depiction of the large scale dynamics of the
system. As an extreme example, an observer whose sole observation was
a high resolution observation along one of the lines of sight which
yields a blue$>$red double peaked asymmetry could be misled into
believing that the source is undergoing systemic infall.

\section{Discussion and Conclusions}

In this study we have concentrated on the implications of the
radiative transfer modelling for our understanding of the chemical
structure of bipolar outflow jet/boundary layers. We have not
attempted to present or test a dynamical model of any
sophistication. This is in contrast with some previous studies
(e.g. Hogerheijde, 1998; Lee et al., 2000; Arce and Goodman, 2002)
which have considered the hydrodynamic activity in the jet, boundary
layer and surrounding cloud in some detail.  Instead, although we do
not constrain the hydrodynamics of the flow, our study makes some
important conclusions concerning the origin and level of anomalous
chemical activity within boundary layers, and how - using a
state-of-the-art radiative transfer code - high resolution
observations can be used to diagnose that activity.

In his study, Hogerheijde (1998) utilized an axisymmetric non-LTE
Monte-Carlo to study a boundary layer whose thickness is 1/5 of the
total width of the (evacuated) jet cavity. His model of the boundary
layer was based on a plausible Couette flow in which the density, flow
velocity and temperature vary across the boundary layer in simple
monotonic fashions, maintaining pressure balance at all points
(Stahler, 1994). Whilst somewhat more sophisticated than our uniform
density flow, the model had problems in reproducing the observed
morphology of L1527 - and in particular the density contrast between
the limb-brightened edges and the body of the outflow cavity.  Indeed,
reasonable fits to the observed morphology - the line intensities and
spatial contrasts in L1527 - were only obtained if the HCO$^+$
abundance in the boundary layer is {\em not} enhanced over that in the
envelope, but the fractional ionization in the boundary layer is very
high [X(e$^-$)$\sim 10^{-4}$].

We do not reproduce this result, and in particular we do not find that
by raising the HCO$^+$ abundance in the boundary layer that the cavity
`fills' due to increased opacity. What we do find is that the density
and contrast are more sensitive to the adopted HCO$^+$ abundance
enhancement and the thickness of the boundary layer, whilst the
detailed morphology is dependent on the assumed outflow shape.  In any
case we do not find that such an extreme level of ionization is
necessary. Indeed, it is not clear what the source of such a high
level of ionization would be. More importantly, it must be remembered
that the dominant loss route of HCO$^+$ in dark clouds is dissociative
recombination, so it would seem reasonable to expect a significant
{\em suppression} of HCO$^+$ if X(e$^-$)$\sim 10^{-5}-10^{-4}$.

Our study is more restricted to a chemical analysis of the interface
and thus adopts a much simpler dynamical model.  Our modelling has
attempted to reproduce the cross-shaped emission seen in L1527 (and
other sources) with the appropriate intensity contrasts in the
outflow lobes. The cause of the brightness in the cross-arms is
essentially due to limb-brightening effects. Clearly, if the boundary
layer is too thick then significant emission will be apparent along
all lines of sight which pass through the boundary layer, and the
morphology would tend to look more like a `bow tie' rather than a
hollow cross.

Our {\em tanh} geometry is somewhat arbitrary, but was chosen because
it matches well the observed morphology. The level of HCO$^+$
enhancement and our plausible explanation of its correlation with the
shape of the outflow then yields patterns of intensity that closely
resemble the high resolution maps; in particular the strong
enhancements (and the location of the peaks of emission) in the
limb-brightened wings, a natural explanation for the limited extent of
the emission and the lack of emission close to the origin.

The essential findings of this paper therefore are:
\begin{enumerate}
\item A strong enhancement of the HCO$^+$ abundance is required in the boundary 
layer between the outflow jet and the surrounding molecular core in order to 
explain the observations of certain outflow sources associated with 
star-forming cores
\item We have proposed a plausible, if simple, mechanism by which this 
enhancement can occur with the observed morphologies and contrasts; shock 
liberation and photoprocessing of molecular material stored in icy mantles. The 
degree of shock activity is closely related to the morphology of the source
\item We have shown how asymmetric, double-peaked line profiles can be 
generated with strong spatial variations in the relative strength of the red 
and blue wings.
\end{enumerate}

\section*{Acknowledgements}
We thank the referee, M. Hogerheijde, for a very constructive report
that helped improve the paper. Some of the calculations were performed
at the HiPerSPACE centre of UCL, which is partially funded by the UK
Joint Research Equipment Initiative. MPR is supported by PPARC. DAW is
supported by the Leverhulme Trust through an Emeritus Fellowship. This
work was partially supported by the National Science Foundation
through a grant for the Institute for Theoretical Atomic, Molecular
and Optical Physics at Harvard University and Smithsonian
Astrophysical Observatory.

\label{lastpage} 

\end{document}